**412**

# Coercivity weighted Langevin magnetisation; A new approach to interpret superparamagnetic and nonsuperparamagnetic behaviour in single domain magnetic nanoparticles

**Dhanesh Kattipparambil Rajan[a] and Jukka Lekkala[b]**

[a,b]Department of Automation Science and Engineering,
Tampere University of Technology, Tampere, P.O. Box 692, FIN-33101 Finland
E-mail: [a] dhanesh.kr@tut.fi
[b] jukka.lekkala@tut.fi

**Abstract**

Superparamagnetism (SPM) is an attractive material property often appearing in nanoscaled single domain (SD) configurations. However, not all SD particles are superparamagnetic, which depends on a few parameters including material type, temperature, measurement time and magneto crystalline anisotropy. The non-linear magnetisation response of magnetic particles can be interpreted by classical Langevinapproach but its applicability is limited to pure SD-SPM behaviour. The classical Langevin equation lacks parameters to account for possible remanence and coercivity in SD regime, resultantly, the SD-nonSPM possibility is left untreated. To solve this issue, we propose a new model by including SD coercivity parameters in classical Langevin equations. The new model 1) combines steady or time varying magnetisation dynamics and temperature or particle size dependent coercivity and 2) helps to calculate coercivity compensated magnetisations and susceptibilities directly. The model covers full spectrum of SD diameters and defines the switching between superparamagnetic and non-superparamagnetic states more precisely.

## 1. Introduction

Superparamagnetic particles have been widely utilised in recent years for their applications in biosensors, targeted drug delivery, therapeutic hyperthermia and tomographic imaging [23][24][25][26]. Superparamagnetism (SPM) is often directly interpreted as a material property achieved by scaling the particle volume down to nanoscale dimensions with the formation of single domain (SD) configuration. But in reality, besides the particle volume, a few other parameters including available thermal energy, magneto crystalline anisotropy and measurement period together determines whether the unique magnetic dipole moment fluctuates randomly ending up in classical superparamagnetic behaviour [27][28][29][30]. Therefore depending on the proportion of these parameters, the SD particles might appear as superparamagnetic (SD-SPM) and non-superparamagnetic (SD-nonSPM). Experimentally, all sorts of SPM behaviour of any material particle is often monitored by magnetisation hysteresis plots and conceived susceptibility measurements [31][32]. There exist a few theoretical models to predict the non-linear magnetisation response with high field saturation mostly using the Langevin approach [31][33][34][35][38]. But the Langevin approach is strictly applicable only in pure SD-SPM cases since it never considers the SD-nonSPM formulation. The SPM to non-SPM transition in SD configuration and the SD remanence and coercive force observed in many experiments [28][29][30] also cannot be interpreted by the conventional Langevin approach. Particle samples from most of the vendors are not strictly mono-disperse, so the probability to have the volume dependent SD remanence in room temperature applications and SPM to non-SPM transition in below room temperature applications is high. In this context we propose a new model, developed from the classical Langevin equations, which combines steady or time varying magnetisation dynamics and temperature or particle size dependent coercive force. The new model helps to calculate coercivity compensated DC and component AC magnetisations and susceptibilities directly from particle and suspension medium properties. The new model covers full spectrum of SD diameters and defines the switching between superparamagnetic and non-superparamagnetic states more precisely. The calculations have been carried out using the material properties of the most used magnetic particle materials of magnetite and maghemite.

## 2. The distinctive SD and SPM configurations

### 2.1. Single domain and superparamagnetic radii

In the absence of an external field, the critical diameter for single domain configuration is a function of exchange length $l_{ex}$ as follows [27]

$$d_{SD} = 72 k l_{ex} \qquad (1)$$

where $k$ is the dimensionless hardness parameter. Substituting for $k$ and $l_{ex}$ yields

$$d_{SD} = 72 \sqrt{\frac{K}{\mu_0 M_s^2}} \sqrt{\frac{A}{\mu_0 M_s^2}} \qquad (2)$$

where $K$ is first anisotropy constant, $\mu_o$ vacuum permeability , $M_s$ saturation magnetisation and $A$ exchange stiffness constant. For a given particle, though its diameter is below $d_{SD}$, it need not necessarily be superparamagnetic below a certain transition temperature since the surrounding thermal energy is not sufficient enough to flip the dipole moment randomly inside the domain in the considered observation time. This leads to state the critical diameter $d_{SPM}$[27] for superparamagnetic behaviour as a function of temperature and magneto crystalline anisotropy as follows,

$$d_{SPM} = 2\sqrt[3]{\frac{6k_b T}{K}} \qquad (3)$$

where $k_b$ is Boltzmann's constant $T$ absolute temperature. The $d_{SD}$ and $d_{SPM}$ calculated for magnetite and maghemite spherical particles at 300K using equation (2) and (3) are given in Table. 1. The variation of $d_{SPM}$ with temperature is shown in 0

Table. 1 **Anisotropy and crystalline parameters defining SD and SPM critical diameters at 300K**[27][36]

|  | First anisotropy constant, $K$ (kJ/m$^3$) | Exchange stiffness constant, $A$ (pJ/m) | Saturation magnetisation, $M_s$ (kA/m) | Single domain critical diameter, $d_{SD}$ (nm) | Superparamagnetic critical diameter, $d_{SPM}$ (nm) |
|---|---|---|---|---|---|
| Magnetite | 13.5 | 13.3 | 446 | ~ 103 | ~ 24 |
| Maghemite | 4.6 | 10 | 380 | ~ 85 | ~ 35 |

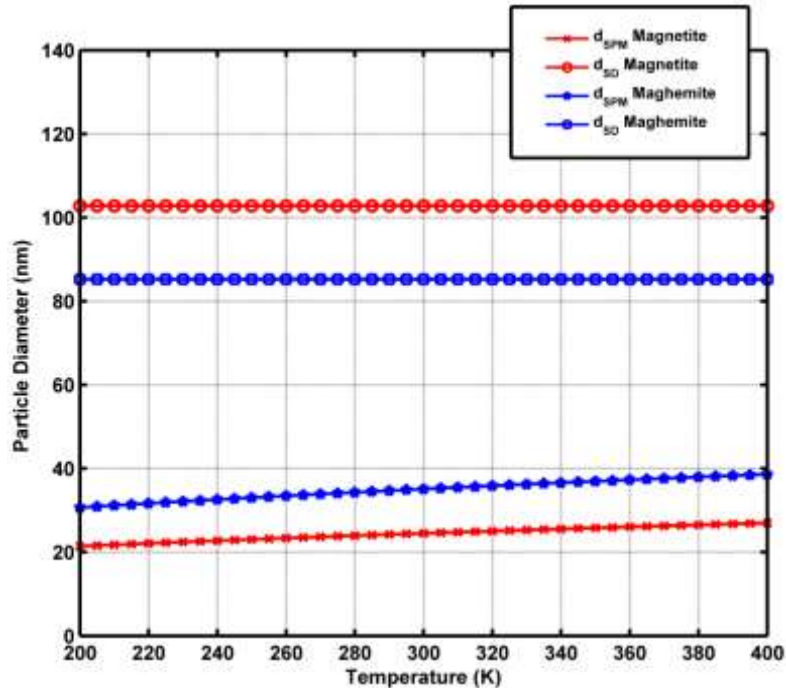

Fig. 1 Single domain critical diameter $d_{SD}$, superparamagnetic diameter $d_{SPM}$ as a function of temperature for magnetite and maghemite particles

## 2.2. Relaxometric parameters and complex susceptibility.

The magnetic moment flips between parallel or antiparallel easy axes and the effective relaxation time constant for a magnetic particle suspension is

$$\tau_{eff} = \frac{\tau_N \tau_B}{\tau_N + \tau_B} \qquad (4)$$

where $\tau_N = \tau_0 \exp(KV/k_bT)$, the Neel relaxation time by Neel-Arrhenius formulation [35] and $\tau_B = (K_r V \eta / 2 k_b T)$, the Brown relaxation time due to Brownian rotational diffusion of suspended particles in carrier medium. $1/\tau_o$ is attempt frequency characteristic to material, $V$ the particle volume, $K_r$ geometric rotational shape factor and $\eta$ carrier medium viscosity. The $\tau_{eff}$ for magnetite and maghemite spherical particles at different SD diameters is in Fig. 2

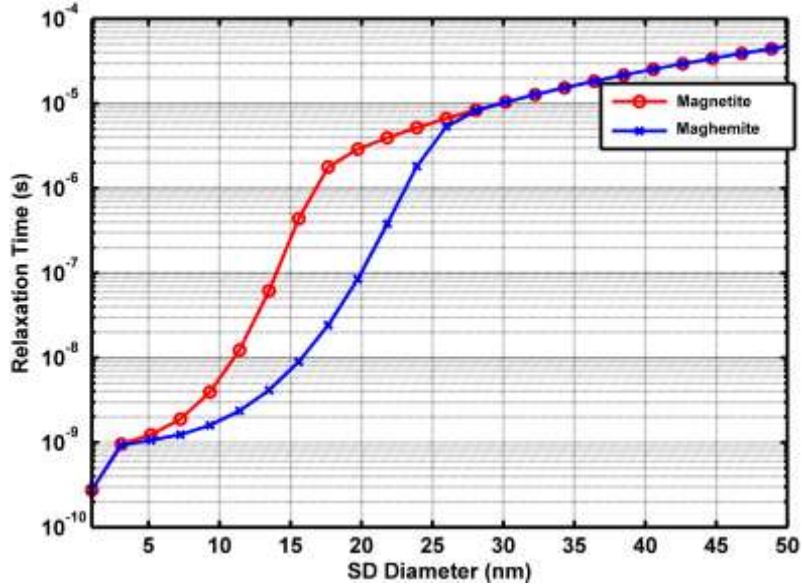

**Fig. 2** **The effective relaxation time $\tau_{eff}$ for magnetite and maghemite spherical particles at different SD diameters**

In an external alternating field, the absolute susceptibility of particle suspension is exclusively determined by the effective relaxation time. The Debye convention to predict the frequency dependent complex susceptibility [31][33] in this case can be written as

$$\chi = \chi' - i\chi'' = \frac{\chi_o}{1+\omega^2 \tau_{eff}^2} - i\frac{\chi_o \omega \tau_{eff}}{1+\omega^2 \tau_{eff}^2} \qquad (5)$$

Where $\chi_o$ is the DC susceptibility and $\omega$ the angular frequency. For equation (5) to be theoretically useful, other approximations for eg. Langevin approximation for $\chi_o$ is essential. For a given volume fraction $\phi$, the Langevin magnetisation [34][35] can be expressed as

50

$$M_{DC} = \phi M_s \left[ \coth(\alpha) - \frac{1}{\alpha} \right] \quad (6)$$

where $\alpha = \pi \mu_o M_s d^3 H_x / 6 k_b T$, $d$ is the particle diameter and $H_x$ the intensity of applied field.

### 2.3. Langevin magnetisation with relaxometric parameters

For an AC field of strength, $H_x \sin \omega t$, the Langevin variable in equation (6) can be modified with the notions $\chi' = \chi_o \cos \omega t$ and $\chi'' = \chi_o \sin \omega t$, to include the real and imaginary susceptibility and frequency components as [35][37][38],

$$M_{AC} = \phi M_s \left[ \frac{1}{1+\omega^2 \tau_{eff}^2} \left( \coth(\alpha \cos \omega t) - \frac{1}{\alpha \cos \omega t} \right) + \frac{\omega \tau_{eff}}{1+\omega^2 \tau_{eff}^2} \left( \coth(\alpha \sin \omega t) - \frac{1}{\alpha \sin \omega t} \right) \right] \quad (7)$$

At 0Hz Equation (7) converges to Equation (6). This equation is useful for predicting volume magnetisation at high temperature and only in the SD-SPM regime and never predicts coercivity or remanence observed in many SD magnetisation experiments [28][29][30].

### 2.4. Langevin magnetisation with relaxometric and coercivity parameters

The temperature dependent SD magnetic coercivity for a randomly oriented non interacting particle system can be expressed as,

$$H_c = H_{co} \left[ 1 - (T/T_B)^{\frac{1}{2}} \right] \quad (8)$$

Where $H_{co} = 2K/\mu_o M_s$ is the coercivity at $0K$ according to the Stoner–Wohlfarth theory [39] & $T_B = KV/k_b \ln(\tau_m/\tau_o)$, is the critical superparamagnetic transition temperature (blocking temperature) [36][40]. By substituting for $H_c$ [41] and $T_B$, the volume dependence of coercivity is derived

$$H_c = H_{co} \left[ 1 - \left( \frac{k_b T}{KV} \ln\left(\frac{\tau_m}{\tau_o}\right) \right)^{\frac{1}{2}} \right] \quad (9)$$

where $1/\tau_m$ is measurement frequency. Equation (8) is valid when $T<T_B$ since $H_c$ cannot have negative values in forward magnetisation. When substituted for $T_B$ in equation (9), the same approximation is followed hence the coercivity $H_c \geq 0$. The temperature and frequency dependence of coercivity of magnetite particles at different single domain diameters is plotted in Fig. 3

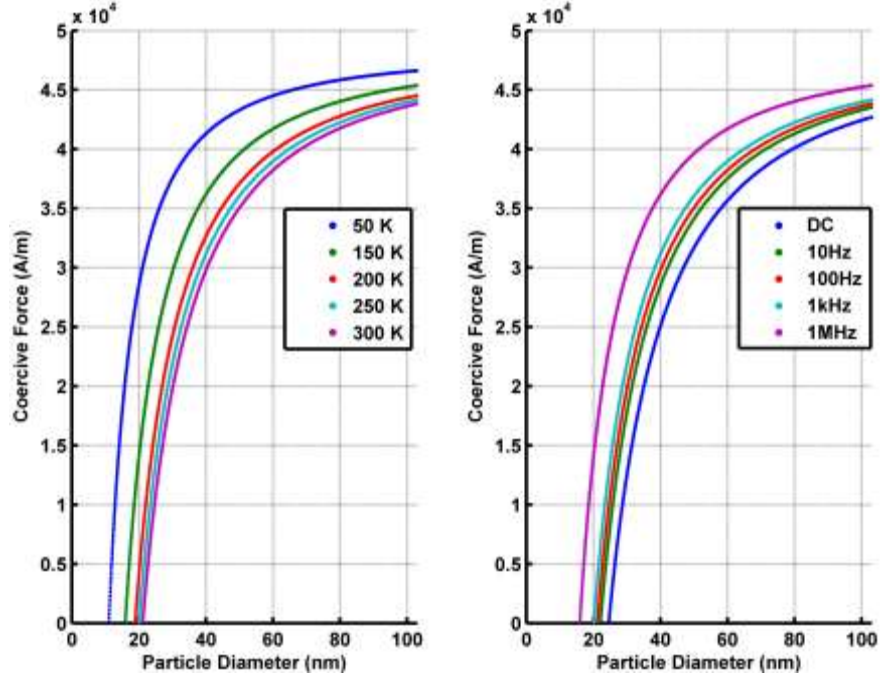

**Fig. 3    Coercivity as a function of particle diameter a) at different temperatures and b) at different field frequencies. The zero coercivity corresponds to the superparamagnetic transition which is clearly a function of temperature (blocking temperature) and measurement frequency.**

To account for coercive force in magnetisation, equation (7) can be modified by including $H_c$ in $\alpha$ and is rewritten for forward and backward measurements as

$$\alpha_{eff} = \frac{\pi \mu_0 M_s d^3 \left( H_x \pm H_c \right)}{6 k_b T} \tag{10}$$

$$M_{AC} = \phi M_s \left[ \frac{1}{1+\omega^2 \tau_{eff}^2} \left( \coth(\alpha_{eff} \cos \omega t) - \frac{1}{\alpha_{eff} \cos \omega t} \right) + \frac{\omega \tau_{eff}}{1+\omega^2 \tau_{eff}^2} \left( \coth(\alpha_{eff} \sin \omega t) - \frac{1}{\alpha_{eff} \sin \omega t} \right) \right]$$

(11)

Equation (11) accounts for the frequency dependent volume magnetisation and volume and temperature dependent coercive force. The equation covers all diameters (SPM and nonSPM) in the complete SD regime. The $M_{AC}$ plots using equation (11) for SD magnetite and maghemite particles at different temperatures are shown in Fig. 4.

The equation for instantaneous volume susceptibility can be derived by differentiating equation (11) with respect to effective field either for forward $H_{eff} = H_x + H_c$    or backward $H_{eff} = H_x - H_c$ magnetisation measurement as follows

$$\chi_{inst} = \frac{d}{dH_{eff}}(M_{AC}) = \frac{-\phi M_s}{1+\omega^2\tau_{eff}^2}\left[\frac{k_1}{H_{eff}}\left(\coth^2(k_1)-1\right)+\frac{\omega\tau_{eff}k_2}{H_{eff}}\left(\coth^2(k_2)-1\right)-\frac{k_2+k_1\omega\tau_{eff}}{k_1k_2H_{eff}}\right]$$

(12)

Where $k_1 = \alpha_{eff}\cos\omega t$ and $k_2 = \alpha_{eff}\sin\omega t$. The $\chi_{inst}$ plots for magnetite and maghemite based on equation (12) are given in Fig. 5.

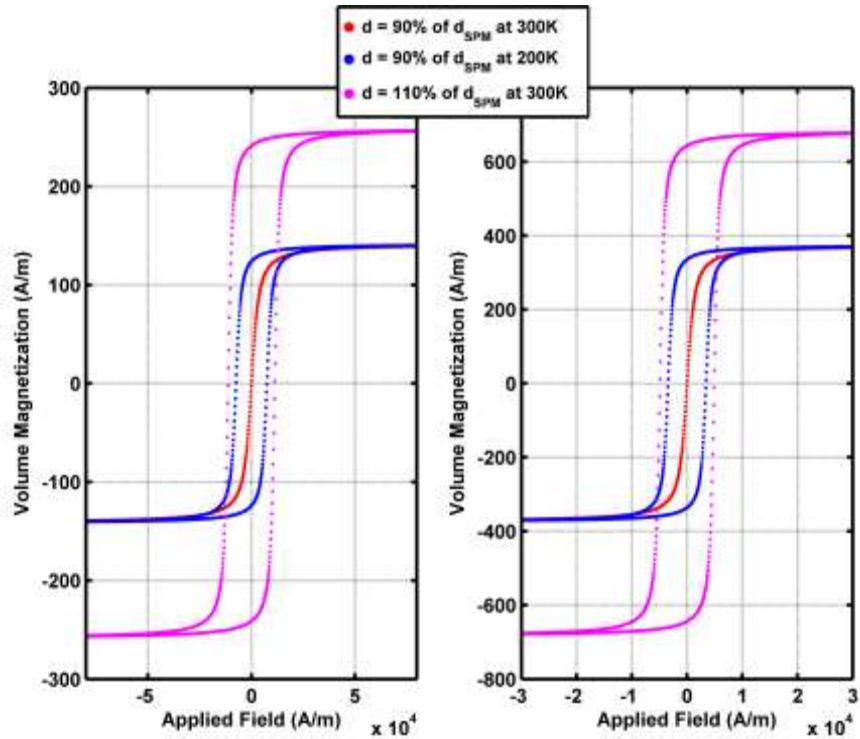

**Fig. 4    The magnetisation plots for a) SD magnetite and b) SD maghemite particles at different temperatures. Two diameters 10% above and below the critical $d_{SPM}$ are considered. For diameters above the $d_{SPM}$ large coercivity appears. Also a superparamagnetic particle at room temperature is not superparamagnetic at a lower temperature. (For computations, f = 10Hz, particle concentration = 0.1mmol/L, suspension medium = distilled water)**

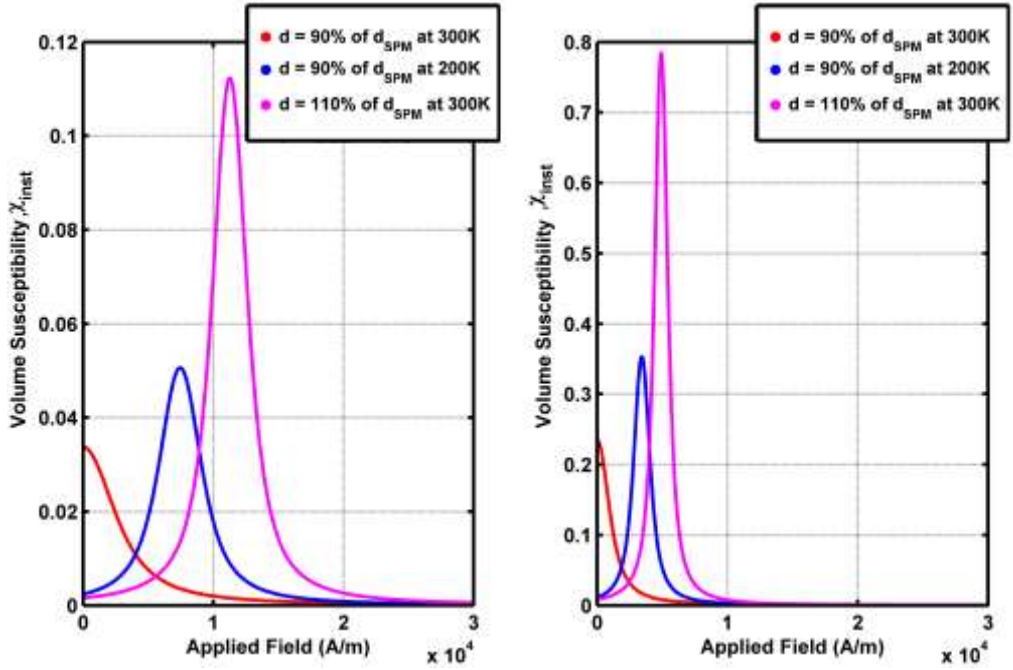

**Fig. 5** The instantaneous susceptibility (full volume susceptibility) plots for two diameters 10% above and below the critical $d_{SPM}$ for a) SD magnetite and b) SD maghemite at different temperatures. The maximal influence of coercive field at low temperature (blue) and above critical $d_{SPM}$ (magenta) is seen as peaks in full susceptibility measurement. As the strength of the applied field increases, the peak susceptibility is seen when the maximum magnetic energy is used to overcome the demagnetising coercive field. Thereafter the superparamagnetic behaviour dominates.

A very useful application of equation (12) is to approximate the DC susceptibility (0Hz) which can be reduced to,

$$\chi_{DC} = \phi M_s \left[ \frac{1}{\alpha_{eff} H_{eff}} - \frac{\alpha_{eff}}{H_{eff}} \left( \coth^2(\alpha_{eff}) - 1 \right) \right] \quad (13)$$

In reality, equation (13) consists of real and imaginary components which can be separately redefined as

$$\chi' = \frac{\phi M_s}{1 + w^2 \tau_{eff}^2} \left[ \frac{1}{\alpha_{eff} H_{eff}} - \frac{\alpha_{eff}}{H_{eff}} \left( \coth^2(\alpha_{eff}) - 1 \right) \right] \quad (14)$$

$$\chi'' = \frac{w \tau \phi M_s}{1 + w^2 \tau_{eff}^2} \left[ \frac{1}{\alpha_{eff} H_{eff}} - \frac{\alpha_{eff}}{H_{eff}} \left( \coth^2(\alpha_{eff}) - 1 \right) \right] \quad (15)$$

The $\chi'$ and $\chi''$ plots for SD- SPM and SD- nonSPM particles for magnetite and maghemite at different frequencies are given in Fig. 6.

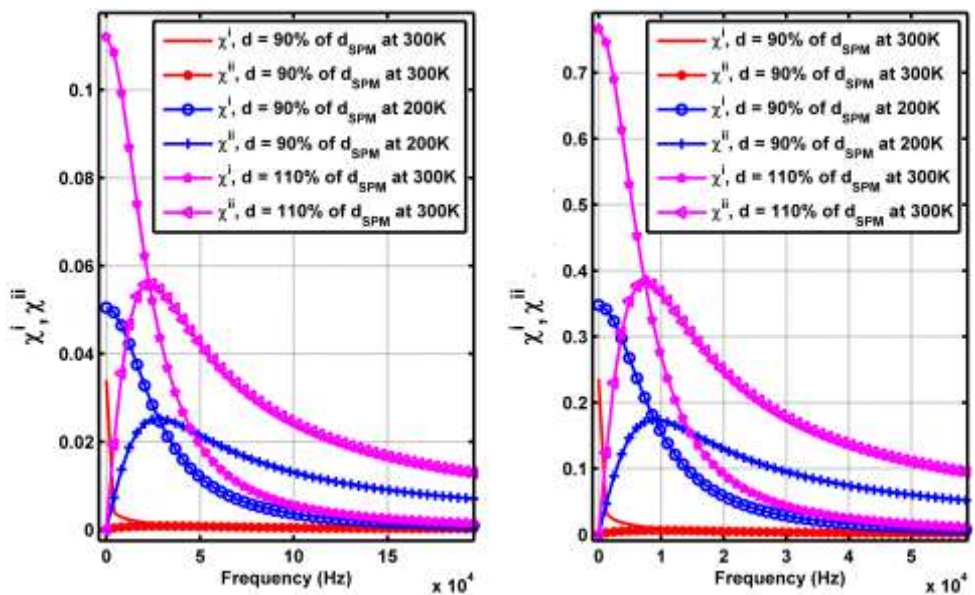

**Fig. 6** The $\chi'$ and $\chi''$ plots for SD- SPM and SD- nonSPM particles for magnetite and maghemite at different frequencies.

Finally the cusp observed in experimental $\chi'$ versus $T$ plots [42] can be effectively predicted by our model as in Fig. 7

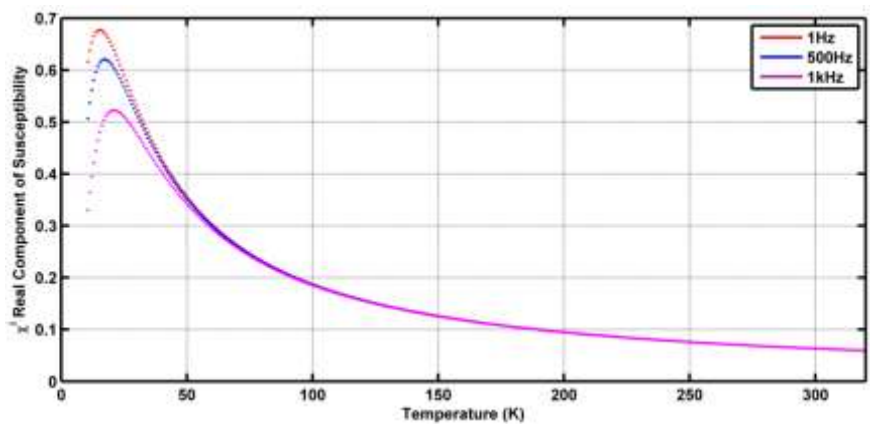

**Fig. 7** $\chi'$ versus $T$ curve for magnetite particle of diameter equals 90% of $d_{SPM}$

### 3. Conclusion

A new model to interpret superparamagnetic and nonsuperparamagnetic behaviour in single domain magnetic nanoparticles weighted by coercivity influence is presented. Equations for directly computing coercivity weighted stationary or time varying magnetisation and susceptibility for non-interacting nanoparticle samples are derived. All equations are derived for monodisperse particles but in reality most of the particle samples from different vendors are polydisperse. The polydispersity can be included in the presented model by replacing the volume fraction '$\phi$' by the *'log normal diameter distribution'* of particles. Direct calculation of magnetisation and susceptibility would be helpful in many biomedical areas where parameters like magnetisation dependent voltage, magnetisation dependent polarisation, magneto optic effect etc. are to be estimated.

## 4. Acknowledgements

The authors acknowledge Hugues de Crémiers, Merck Millipore for certain helpful discussions. This work was carried out during the tenure of funding from the project MPI-SPARE of Human Spare Parts Project supported by Finnish Funding Agency for Technology and Innovation (TEKES) and the Council of Tampere Region